\documentclass[twocolumn,secnumarabic,amssymb, amsmath, nobibnotes, aps, prb, citeautoscript, showpacs, reprint]{revtex4-1}

\usepackage{graphicx}
\usepackage{dcolumn}
\usepackage{bm}
\usepackage{color} 

\begin{document}

\title{Real-space renormalized dynamical mean field theory}

\author{Dai Kubota,$^{1}$ Shiro Sakai,$^{2}$ and Masatoshi Imada$^{1}$}
\affiliation{${}^{1}$Department of Applied Physics, University of Tokyo, Hongo, Tokyo 113-8656, Japan. \\
${}^{2}$Center for Emergent Matter Science, RIKEN, Hirosawa, Wako, Saitama 351-0198, Japan.}



\begin{abstract}
We propose {\it real-space renormalized dynamical mean field theory} (rr-DMFT) to deal with large clusters in the framework of a cluster extension of the DMFT.
In the rr-DMFT, large clusters are decomposed into multiple smaller clusters through a real-space renormalization.
In this work, the renormalization effect is taken into account only at the lowest order with respect to the intercluster coupling, which nonetheless reproduces exactly both the noninteracting and atomic limits.
Our method allows us large cluster-size calculations which are intractable with the conventional cluster extensions of the DMFT with impurity solvers, such as the continuous-time quantum Monte Carlo and exact diagonalization methods.
We benchmark the rr-DMFT for the two-dimensional Hubbard model on a square lattice at and away from half filling, where the spatial correlations play important roles.
Our results on the spin structure factor indicate that the growth of the antiferromagnetic spin correlation is taken into account beyond the decomposed cluster size.
We also show that the self-energy obtained from the large-cluster solver is reproduced by our method better than the solution obtained directly for the smaller cluster.
When applied to the Mott metal-insulator transition, the rr-DMFT is able to reproduce the reduced critical value for the Coulomb interaction comparable to the large cluster result.

\end{abstract}

\pacs{71.10.Fd, 71.27.+a, 71.30.+h, 75.10.-b} 

\maketitle

\section{INTRODUCTION}\label{sec:intro}
Strongly correlated electron systems have intensively been studied in the past decades.
Various numerical methods have been developed to study correlated-electron models such as the Hubbard model.
Nevertheless, the complete understanding of them remains open \cite{RevModPhys.70.1039}. 
Among the methods, the dynamical mean field theory (DMFT) \cite{PhysRevLett.62.324, RevModPhys.68.13} has an advantage in calculating dynamical properties, by taking into account full temporal fluctuations.
Although its original single-site formalism completely ignores spatial correlation effects, there are several approaches to take them into account.
The approaches include cluster extensions of the DMFT, such as the cellular dynamical mean field theory (CDMFT) \cite{PhysRevLett.87.186401}, dynamical cluster approximation (DCA) \cite{PhysRevB.58.R7475}, and variational cluster approximation \cite{PhysRevLett.91.206402}, which incorporate the spatial correlation effects within the cluster.
Alternatively, diagrammatic extensions \cite{doi:10.1143/JPSJ.75.054713, PhysRevB.75.045118, PhysRevB.77.033101} have been proposed.

In these approaches, the original lattice model is mapped onto an effective quantum impurity model consisting of interacting sites and a noninteracting fermionic bath. 
The effective impurity model is solved, for example, by the continuous-time quantum Monte Carlo (CT-QMC) \cite{PhysRevB.72.035122, 0295-5075-82-5-57003, PhysRevLett.97.076405}, pioneered by Beard and Wiese {\it et al.} \cite{PhysRevLett.77.5130, JETP.87.2.310}, and exact diagonalization (ED) \cite{PhysRevLett.72.1545} methods.
In these cluster extensions of the DMFT, the range of the spatial correlations incorporated into the numerical solution is severely limited by the cluster size.
In fact, with the increase in the cluster size, the computational cost rapidly increases, which easily makes the practical computation intractable.

The computational cost of the CT-QMC depends on its algorithm: In the weak-coupling algorithms \cite{PhysRevB.72.035122, PhysRevLett.82.4155, 0295-5075-82-5-57003}, the cost scales as $O[\left( N_{c}\beta U\right)^{3}]$ in the absence of the negative-sign problem (while a more efficient algorithm has been recently proposed \cite{PhysRevB.91.241118}), when we apply an $N_{c}$-site cluster-DMFT to the Hubbard model with the onsite repulsive interaction $U$ at the inverse temperature $\beta$.
In another algorithm based on the hybridization expansion \cite{PhysRevLett.97.076405}, its cost grows as $O[\beta^{3}\exp{\left( N_{c}\right)}]$ with increasing $N_{c}$.
On the other hand, in the ED, the free fermion bath of the quantum cluster model is represented by a finite number ($N_b$) of fermions.
The numerical cost then exponentially increases as $O[4^{(N_{c}+N_{b})}]$.
Thus, the accessible cluster size is severely limited for all these solvers.
Furthermore, this difficulty becomes more serious when we consider the orbital degrees of freedom \cite{PhysRevLett.94.026404, PhysRevLett.104.026402, PhysRevB.89.195146, PhysRevB.91.235107}.

The dilemma between the accuracy and the computational cost has led several authors to propose methods to obtain better results from small clusters.
These include improved interpolation schemes in the DCA \cite{PhysRevB.93.165144, PhysRevB.88.115101} and combinations of the cluster and diagrammatic extensions of the DMFT \cite{PhysRevB.84.155106, JETP86Hafermann, PhysRevB.91.121111}.

A sufficiently large cluster is required to understand strong correlation effects in low-dimensional systems.
For example, the self-energy has a strong momentum dependence in doped Mott insulators \cite{PhysRevLett.84.522, PhysRevB.66.075102, PhysRevLett.92.126401, PhysRevLett.95.106402, kuchinskii2005destruction, PhysRevB.73.165114, PhysRevB.74.125110, PhysRevB.76.104509, PhysRevB.80.165126,  PhysRevLett.102.056404, PhysRevB.82.134505, PhysRevB.82.155101, PhysRevB.84.075161, PhysRevB.85.035102} in two dimensions.
Continuous phase transitions and regions near quantum criticalities in general require accurate treatments of long-range spatial correlations. 
From a quantitative point of view, small cluster studies of the two-dimensional Hubbard model substantially overestimate the critical interaction $U_{c}$ of the Mott metal-insulator transition (MIT) \cite{RevModPhys.70.1039, PhysRevB.76.045108,PhysRevLett.101.186403} and the magnitude of the order parameter of the $d$-wave superconductivity \cite{PhysRevB.90.115137}.

In this article, we introduce an approach based on the CDMFT combined with the real-space renormalization, which we call {\it real-space renormalized dynamical mean field theory} (rr-DMFT).
In the rr-DMFT, a large cluster model is decomposed into multiple smaller-cluster problems which can be solved at a considerably smaller computational cost than the former model, while the spatial correlations up to the size of the original large cluster are reasonably taken into account through a renormalization of the dynamical mean field for the smaller clusters.

The article is organized as follows:
In Sec.~\ref{sec:formalism}, the rr-DMFT is introduced with several practical examples of the decomposition of a large cluster model into multiple small-size problems.
Section \ref{sec:result} describes benchmark results for the two-dimensional Hubbard model on a square lattice.
The short-range antiferromagnetic correlations play key roles in this system.
We show that our method correctly takes them into account through the results on the spin structure factor and the description of the self-energy beyond the decomposed cluster size.
The results obtained by this method are also favorably compared with the conventional CDMFT results on the MIT, where the reduction of the critical strength of the onsite Coulomb repulsion is properly reproduced \cite{PhysRevB.76.045108, PhysRevLett.101.186403}.
We also calculate and compare the density of states with using the exact diagonalization solver.
Finally, Sec.~\ref{sec:conclusion} is devoted to conclusions.

\section{FORMALISM}\label{sec:formalism}
In the present rr-DMFT approach, we approximate the original large cluster model by mapping it onto that of multiple small clusters, and then solve it by using a quantum impurity solver for small clusters.
In this paper, we employ the lowest-order cumulant expansion to take into account the inter-cluster correlations.

In Sec.~\ref{ssec:ren},we elaborate how we reduce a large cluster model to small cluster problems through a real-space renormalization.
In Sec.~\ref{ssec:alg}, we describe an algorithm to systematically calculate all the elements of the Green's function using the mapping.
It is shown that if an $N$-site impurity model is solvable, the Green's function for $2N$-site cluster models is obtained within our method.
By repeating this procedure, we can solve the large cluster problem only with the impurity solver for the small cluster.

\subsection{Renormalization process}\label{ssec:ren}
We introduce the rr-DMFT for the Hubbard model, defined by the Hamiltonian,
	\begin{equation}
	H=\sum_{i,j,\sigma}t_{ij}c_{i\sigma}^{\dagger}c_{j\sigma}
		-\mu \sum_{i\sigma}c_{i\sigma}^{\dagger}c_{i\sigma}
		+U\sum_{i}n_{i\uparrow}n_{i\downarrow}.
	\end{equation}
Here, $c_{i\sigma} ( c_{i\sigma}^{\dagger})$ annihilates (creates) an electron with spin $\sigma =\uparrow ,\downarrow$ at site $i$.
$n_{i\sigma}=c_{i\sigma}^{\dagger}c_{i\sigma}$ is the density operator.
The hopping amplitude between sites $i$ and $j$ is denoted by $t_{ij}$.
$\mu$ is the chemical potential and $U (>0)$ is the onsite Coulomb repulsion.
The CDMFT maps the Hubbard model onto an effective quantum impurity model whose action is given by
	\begin{eqnarray}
	S_{C}&&=- \int_{0}^{\beta}d\tau \int_{0}^{\beta}d\tau^{\prime}\sum_{i,j\in C, \sigma}c_{i\sigma}^{\dagger}(\tau )\mathcal{G}^{-1}_{0,ij\sigma }(\tau -\tau^{\prime})c_{j\sigma}(\tau^{\prime}) \nonumber \\
 	&&+U\int_{0}^{\beta}d\tau \sum_{i\in C}n_{i\uparrow}(\tau )n_{i\downarrow}(\tau ),
	\end{eqnarray}
where $\beta$ is the inverse temperature and $C$ represents the cluster.
The dynamical one-body part $\hat{\mathcal{G}}^{-1}_{0}(\tau )$ is the Weiss function.
The rr-DMFT also starts with this mapping onto a cluster model and provides an approximate way to calculate the Green's function without using any $C$-size impurity solvers as we describe in the following.

The single-particle Green's function $G_{ij\sigma}(\tau )$ of $S_{C}$ is defined by
	\begin{align}
	G_{ij\sigma}(\tau )\equiv &\frac{\mathrm{Tr}_{C}\left( T_{\tau}c_{i\sigma}(\tau )c_{j\sigma}^{\dagger}(0)e^{-S_{C}}\right)}{\mathrm{Tr}_{C}\left( e^{-S_{C}}\right)}.
	\end{align}
Here, $\mathrm{Tr}_{C}$ means the trace over the degrees of freedom within the cluster $C$, and $T_{\tau}$ is the time-ordering operator.

We define the action $S_{C_{1}}^{\mathrm{exact}}$ for a smaller cluster $C_{1}$ by
	\begin{eqnarray}
	\frac{e^{-S_{C_{1}}^{\mathrm{exact}}}}{\mathrm{Tr}_{C_{1}}e^{-S_{C_{1}}^{\mathrm{exact}}}}= \frac{\mathrm{Tr}_{C_{2}}e^{-S_{C}}}{\mathrm{Tr}_{C}e^{-S_{C}}},
	\end{eqnarray}
where $C_{2}$ is a part of $C$ other than $C_{1}$.
With $S_{C_{1}}^{\mathrm{exact}}$, $G_{ij\sigma}$ ($i,j \in C_{1}$)  is obtained as
	\begin{align}
	G_{ij\sigma}(\tau ) =&\frac{\mathrm{Tr}_{C_{1}}\left( T_{\tau}c_{i\sigma}(\tau )c_{j\sigma}^{\dagger}(0)e^{-S_{C_{1}}^{\mathrm{exact}}}\right)}{\mathrm{Tr}_{C_{1}}\left( e^{-S_{C_{1}}^{\mathrm{exact}}}\right)}.
	\label{eq:Gcalc}
	\end{align}
Here we need only a $C_{1}$-size impurity solver to calculate Eq.~(\ref{eq:Gcalc}).

However, $S_{C_{1}}^{\mathrm{exact}}$ generally contains $n$-body interaction terms ($n=1,2,\dots $) and is not tractable with standard impurity solvers.
We therefore replace $S_{C_{1}}^{\mathrm{exact}}$ with the action $S_{C_{1}}$, which is a simple quantum impurity problem obtained from a perturbative treatment of the coupling between $C_{1}$ and $C_{2}$.
In Sec.~\ref{sec:result}, we examine the validity of this replacement by comparing the rr-DMFT results with the CDMFT fully performed for $C$.
At the lowest order in the cumulant expansion \cite{PhysRevB.43.8549}, the action is written as
	\begin{eqnarray}
	&&S_{C_{1}}^{\mathrm{exact}}\simeq S_{C_{1}} \nonumber \\
	&& \ \ \ \ \ \ =- \int_{0}^{\beta}d\tau \int_{0}^{\beta}d\tau^{\prime}\sum_{i,j\in C_{1}, \sigma}c_{i\sigma}^{\dagger}(\tau )\tilde{\mathcal{G}}^{-1}_{0,ij\sigma}(\tau -\tau^{\prime})c_{j\sigma}(\tau^{\prime}) \nonumber \\
 	&& \ \ \ \ \ \ +U\int_{0}^{\beta}d\tau \sum_{i\in C_{1}}n_{i\uparrow}(\tau )n_{i\downarrow}(\tau ),
	\label{eq:renaction}
	\end{eqnarray}
with
	\begin{eqnarray}
		&&\tilde{\mathcal{G}}_{0,ij\sigma}^{-1}(i\omega_{n})=\mathcal{G}_{0,ij\sigma}^{-1}(i\omega_{n}) \nonumber \\
	&&\ \ \ \ \ \ \ \ \ \ -\sum_{k,l\in C_{2}}\mathcal{G}^{-1}_{0,ik\sigma}(i\omega_{n})G^{(C_{2})}_{kl\sigma}(i\omega_{n})\mathcal{G}^{-1}_{0,lj\sigma}(i\omega_{n}),
	\label{eq:renweiss}
	\end{eqnarray}
where $\omega_{n}=(2n+1)\pi /\beta$ is the Matsubara frequency.
Here, $\hat{G}^{(C_{2})}$ is the Green's function for the action
	\begin{eqnarray}
	S^{(C_{2})}=&&-\int_{0}^{\beta}d\tau \int_{0}^{\beta}d\tau^{\prime}\sum_{i,j\in C_{2},\sigma}c_{i\sigma}^{\dagger}(\tau )\mathcal{G}_{0,ij}^{-1}(\tau -\tau^{\prime})c_{j\sigma}(\tau^{\prime}) \nonumber \\
	&&+U\int_{0}^{\beta}d\tau \sum_{i\in C_{2}}n_{i\uparrow}(\tau )n_{i\downarrow}(\tau ),
	\label{eq:sc2}
	\end{eqnarray}
namely,
	\begin{eqnarray}
	G^{(C_{2})}_{ij\sigma}(\tau )=\frac{\mathrm{Tr}_{C_{2}}\left( T_{\tau}c_{i\sigma}(\tau )c_{j\sigma}^{\dagger}(0)e^{-S^{(C_{2})}}\right)}{\mathrm{Tr}_{C_{2}}\left( e^{-S^{(C_{2})}}\right)}.
	\end{eqnarray}
The second term on the right-hand side of Eq.~(\ref{eq:renweiss}) represents the renormalization effect on the Weiss function, which effectively takes into account the $C_2$ degrees of freedom.
This is similar to the cavity method for finite-size lattices, used in the nano-DMFT \cite{PhysRevLett.99.046402}.
In this approximation, $t/U$ is the basic small parameter.
The rr-DMFT therefore reproduces the exact atomic limit.
Furthermore, the exact noninteracting limit is also recovered.
The size of the boundary between $C_{1}$ and $C_{2}$ also controls the accuracy because the number of dominant terms in the renormalization is determined by the number of couplings between two clusters, which is controlled by the size of the boundary.
The renormalization effect on the interaction term in $S_{C_{1}}$ comes in when we consider higher-order cumulants.
However, the lowest-order renormalization of Eq.~(\ref{eq:renweiss}) gives results in reasonable agreement with the CDMFT results even in the intermediate-coupling regime, as we show in Sec.~\ref{sec:result}.

More specifically, the lowest-order approximation gives a reasonable value for the Green's function in the cluster $C=C_1+C_2$ while it neglects the vertex corrections connecting the clusters $C_{1}$ and $C_{2}$. 
This lack of the intercluster vertex corrections limits the applicability of the present rr-DMFT to the system with enhanced long-range correlations, e.g., in the vicinity of quantum critical points.
The higher-order corrections from the real-space renormalization will mitigate the problem since they include a part of the vertex functions beyond the size of $C_1$.

\begin{figure}[t]
  \begin{center}
   \includegraphics[width=0.9\columnwidth]{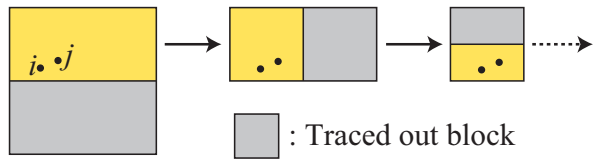}
  \end{center}
  \caption{(Color online) 
  Scheme of step-by-step renormalization to compute the Green’s function $G_{ij}$ in rr-DMFT.
  The large square in the left panel represents the original large cluster, and $i$ and $j$ represent the sites in the cluster.
  Here, the gray regions, which does not include $i$ and $j$, are traced out, leaving the yellow area, which is divided into two regions in the next step. 
  }
  \label{fig:examplerenorm}
\end{figure}
In this subsection, the action $S_{C}$ is decomposed into two cluster problems described by $S_{C_{1}}$ and $S^{(C_{2})}$.
This means that the Green's function for $C_{1}$ is approximately obtained by using not $C$-size but $C_{1}$- and $C_{2}$-size impurity solvers only.
Here, $C_{1}$ and/or $C_{2}$ can still be large for impurity solvers, i.e., beyond the size tractable with the direct use of the impurity solvers.
In such a case, we iterate the real-space renormalization step by step to trace out further degrees of freedom, as illustrated in Fig.~\ref{fig:examplerenorm}.
This figure exemplifies the calculation of the Green's function $G_{ij}$.
We implement such a renormalization process for any site pairs ($i,j$) in the cluster $C$ to obtain all the elements of the Green's function.
In the next subsection, we describe a practical algorithm for this.

\subsection{Algorithm to calculate Green's function}\label{ssec:alg}
For the sake of simplicity, we restrict ourselves to two-dimensional systems and solve an $N_{c}=2^{n}N_{s}$ cluster with an impurity solver for $N_{s}=2$ sites.
It is straightforward to extend the algorithm to arbitrary $N_{s}$, clusters and dimensions.

First, we consider the case of $n=1$.
The Green's function for the $N_{c}=4$ sites is obtained as follows:
We first choose two sites, $l$ and $m$, from the cluster and trace out the other sites, according to the procedure described in Sec.~\ref{ssec:ren}.
We then obtain an action for the two sites $l$ and $m$.
The Green's function $G_{lm}$ is calculated from this action with the two-site impurity solver.
By iterating such a calculation for all $(l,m)$ pairs in the $N_c$-site cluster, we obtain all the elements of the Green's function.

\begin{figure}[tb]
  \begin{center}
   \includegraphics[width=0.9\columnwidth]{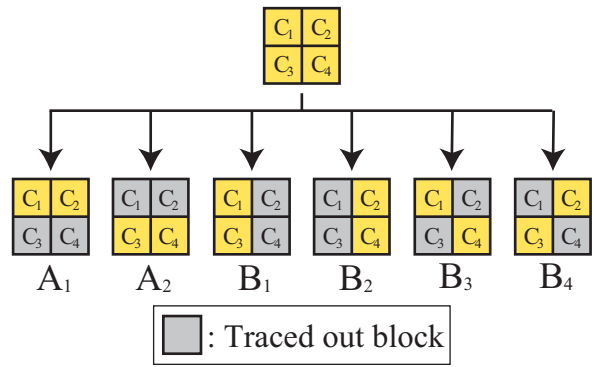}
  \end{center}
  \caption{(Color online) 
  Decomposition of $2^{n+1}$-site cluster into four blocks $C_{i} (i=1,\dots ,4)$.
  Here, $A_{k} (k=1,2)$ and $B_{k} (k=1,\dots ,4)$ denote the groups described in the text.
  Each $C_{i}$ consists of $2^{n-1}$ sites.
  }
  \label{fig:rrdmftalgorithm1}
\end{figure}
We proceed to a general case.
Suppose that we can solve a $2^{n}$-site problem and would like to solve an impurity model of $2^{n+1}$ sites.
We decompose the latter model into several $2^{n}$-site problems (Fig.~\ref{fig:rrdmftalgorithm1}) as follows:
We break up $2^{n}N_{s}=2^{n+1}$-site $C$ into four blocks, $C_{i}$ ($i=1,\dots ,4$), where each $C_{i}$ consists of $2^{n-1}$ sites. 
The elements of the Green's function, $G_{ij}$, are categorized into six groups:
	\begin{itemize}
	\item $A_{1}$= $\{ G_{C_{1}C_{1}}, G_{C_{1}C_{2}}, G_{C_{2}C_{1}}, G_{C_{2}C_{2}} \}$ ,
	\item $A_{2}$= $\{ G_{C_{3}C_{3}}, G_{C_{3}C_{4}}, G_{C_{4}C_{3}}, G_{C_{4}C_{4}} \}$ ,
	\item $B_{1}$= $\{ G_{C_{1}C_{3}}, G_{C_{3}C_{1}} \}$ ,
	\item $B_{2}$= $\{ G_{C_{2}C_{4}}, G_{C_{4}C_{2}} \}$ ,
	\item $B_{3}$= $\{ G_{C_{1}C_{4}}, G_{C_{4}C_{1}} \}$ ,
	\item $B_{4}$= $\{ G_{C_{2}C_{3}}, G_{C_{3}C_{2}} \}$ .
	\end{itemize}
Here, we defined $G_{C_{l}C_{m}}$ as a set $\{ G_{ij} | i\in C_{l}\wedge j\in C_{m} \}$, and omit the spin index $\sigma$ for brevity.
The idea behind this categorization is as follows.
If we naively categorize $G_{ij}$ elements into six groups, such as 
$A_{k}^{\prime}=\{ G_{C_{i}C_{i}}, G_{C_{i}C_{j}}, G_{C_{j}C_{i}}, G_{C_{j}C_{j}} \}$ for $i,j=1,\dots ,4$ and $i\neq j$, they have common elements.
For example, $\{ G_{C_{1}C_{1}},G_{C_{1}C_{2}},G_{C_{2}C_{1}},G_{C_{2}C_{2}} \}$ and $\{ G_{C_{1}C_{1}},G_{C_{1}C_{3}},G_{C_{3}C_{1}},G_{C_{3}C_{3}} \}$ share $G_{C_{1}C_{1}}$. 
In order to reduce the total computational cost, we have removed such overlaps in the above classification: 
Namely, we have excluded the diagonal elements, such as $G_{C_{l}C_{l}}$, from the $B_k$ groups.
We have thus decomposed a problem of calculating $G_{ij}$ for $i,j \in C$ into six tasks of calculating the $A_{k} (k=1,2)$ and $B_{k} (k=1,\dots ,4)$ elements, which include and exclude diagonal elements, respectively.

To obtain each element of the Green's function, we first need to construct an effective action $S_{ij}$ consisting of $C_{i}$ and $C_{j}$ degrees of freedom through the real-space renormalization process described in Sec.~\ref{ssec:ren}.
Since $S_{ij}$ consists of only $2^n$ sites, it is solvable by the assumption.
For example, for the calculation of the $A_1$ element, we construct $S_{12}$ and solve it.

In constructing $S_{ij}$, we need to use the $N_{s}$-site impurity solver $2^{n}/N_{s}$ times to eliminate $2^{n}$-site degrees of freedom in the $2^{n+1}$-site problem.
In practice, we first choose $N_{s}$ sites from the $2^{n}$ sites and trace out the former by the real-space renormalization.
Next we again choose $N_{s}$ sites from the rest $(2^{n}-N_s)$ sites and trace out the former.
By iterating such processes $2^{n}/N_{s}$ times, we can trace out $2^{n}$ sites.

In general, there are many different ways to choose $N_{s}$ sites from the $2^n$ sites.
In practice, we choose a way to make the solved $N_s$-site cluster as compact as possible since correlations between neighboring sites are more important than those between distant sites.
In fact, the short-range correlations play key roles in the pseudogap regime \cite{PhysRevB.73.165114, PhysRevB.74.125110, PhysRevLett.102.056404, PhysRevB.82.134505, PhysRevB.85.035102}, the MIT \cite{PhysRevB.76.045108, PhysRevLett.101.186403} and the $d$-wave superconductivity transition \cite{PhysRevLett.95.237001}.
As we remarked in the previous subsection, this is also justified by the following discussion: The renormalization terms at any orders always include the coupling terms between decomposed clusters [see Eq.~(\ref{eq:renweiss})], and therefore its magnitude depends on the number of the coupling. This number generally becomes small when we decompose the cluster as compact as possible.
We describe a detailed algorithm in Appendix B and demonstrate that the results are not significantly affected by the ways of tracing out as long as we use compact clusters.

In order to obtain all the elements of the Green's function, $G_{ij}$, we need to make $i$ and $j$ sweep over the $N_c$-site cluster. 
Here, it is useful to note that the number of times to use the $N_{s}$-site impurity solver can be minimized by sharing, as much as possible, the action of the sites which are traced out. 
For example, we consider tracing out $C_{1}$-degrees of freedom in the cluster $C=C_{1}+C_{2}+C_{3}+C_{4}$. 
The resultant action consisting of the $C_{2}+C_{3}+C_{4}$ degrees of freedom can be commonly used to derive actions, $S_{23}$, $S_{34}$, and $S_{24}$.
Thus, by sharing the action appearing in each intermediate step as much as possible, we can save the number of times invoking the impurity solver and make the algorithm more efficient.
This enables us to solve the $N_{c}$-site quantum impurity model by using the $N_{s}$-site impurity solver $O\left( N_{c}/N_{s}\right)^{2}$ times (see Appendix A).

In addition, it is possible to further reduce the number using the impurity solver if we utilize the symmetries of the cluster.
For instance, we can use the $C_{4v}$ symmetry for an $N_{c}=L\times L$ square cluster unless a symmetry-breaking order occurs.

\section{RESULTS}\label{sec:result}
In this section, we apply the rr-DMFT to the two-dimensional Hubbard model on the square lattice.
First, we compare the computational cost of the rr-DMFT with that of the CDMFT.
Second, we calculate several physical quantities, such as the spin structure factor, self-energy and density of states, to examine the accuracy of the rr-DMFT.
We also study the MIT phase boundary and show that our method offers a good approximation to the CDMFT results.
We use the continuous-time auxiliary-field quantum Monte Carlo method (CT-AUX) and, in addition, the ED in Sec.~\ref{ssec:cost} and \ref{ssec:dos}, for solving the impurity problem.
Through this section, although the rr-DMFT is of course more advantageous when the regular CDMFT is not tractable any more, we basically restrict ourselves to cases where the regular CDMFT is feasible to allow the comparison with our method as benchmarks.

\subsection{Computational cost of rr-DMFT}\label{ssec:cost}
As we described in the introduction, the computational cost of the CT-QMC and ED increases rapidly with the cluster size.
Since the rr-DMFT uses only an $N_{s}$-site impurity solver, even though we need to use it $O[\left( N_{c}/N_{s}\right)^{2}]$ times (see Appendix A), the rr-DMFT largely reduces the total computational cost.
In the case of the CT-AUX, the computational cost is reduced from $O[N_{c}^{3}]$  to $O[N_{c}^{2}N_{s}]$ even in the absence of the negative-sign problem.
Furthermore, the rr-DMFT mitigates the sign problem.
In the case of the ED, the cost is reduced from $O[4^{N_{c}(n_{b}+1)}]$ to $O[4^{N_{s}(n_{b}+1)}\left( N_{c}/N_{s}\right)^{2}]$, where $n_{b}\equiv N_{b}/N_{c}$.

\begin{figure}[tb]
  \begin{center}
   \includegraphics[width=\columnwidth]{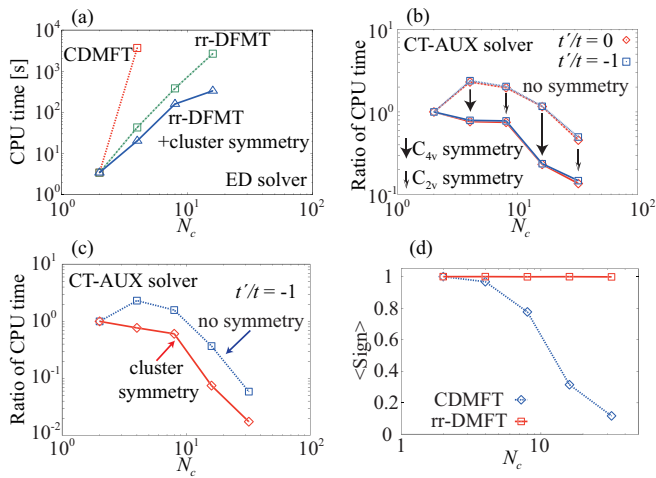}
  \end{center}
  \caption{ (Color online) 
  CPU time spent in solving quantum impurity models of $N_{c}=2\times 1, 2\times 2, 4\times 2, 4\times 4$ and $8\times 4$ at a single core (Xeon X 5690 3.47GHz).
  We fix $N_{s}=2$ for the rr-DMFT and take $U/t=8$ at half filling.
  (a) Comparison with using ED as the impurity solver for $n_{b}=2$ at $T=0$ and $t^{\prime}/t=0$. 
  (b) Ratio of CPU time, namely CPU time of rr-DMFT scaled by that of CDMFT, for CT-AUX solver. 
  Here we take 10000 Monte Carlo samples, and show the results for $t^{\prime}/t=0$ (red solid and dashed lines with diamonds) and $-1$ (blue solid and dashed lines with squares) at $\beta t=4$ with and without using the cluster symmetry ($C_{4v}$ for $2\times 2$ and $4\times 4$ clusters and $C_{2v}$ for $2\times 1$, $4\times 2$ and $8\times 4$ clusters).
  Solid (dashed) lines represent the results with (without) utilizing the cluster symmetry.
  In contrast to the rr-DMFT, the use of the cluster symmetry does not affect much the total computational cost in the CDMFT. 
  This is because in the CDMFT it is always necessary to solve the whole impurity problem while the symmetry can reduce the number of the solved impurity problems in the rr-DMFT.
  (c) Results of (b) for $t^{\prime}/t=-1$ normalized by the average sign shown in (d).
   }
  \label{fig:cputime}
\end{figure}
Figure \ref{fig:cputime} compares the CPU time (measured with a single processor) for solving an $N_{c}$-site quantum impurity problem for the two-dimensional half-filled Hubbard model at $U/t=8$.
We use the ED at $T=0$ and CT-AUX with the submatrix updates \cite{PhysRevB.83.075122} as impurity solvers, and set $N_{s}=2$ in all the rr-DMFT simulations.

In the ED, we use $n_{b}=2$ to represent the Weiss function and solve the problem with the Lanczos method following Ref.~\onlinecite{PhysRevLett.72.1545}.
In Fig.~\ref{fig:cputime}(a), we measure the CPU time of the CDMFT and rr-DMFT for $t^{\prime}/t=0$ at $T=0$.
For the latter, we show two results with and without using the cluster symmetry: 
We consider the $C_{4v}$ symmetry for $L\times L$ clusters, and $C_{2v}$ symmetry for $L\times L^{\prime}$ clusters ($L\neq L^{\prime}$).
The results show that the numerical cost of the rr-DMFT increases only in a power of $N_{c}$ and that it is further reduced by the use of the cluster symmetry.

Next we examine the case of the CT-AUX solver for $t^{\prime}/t=0$ and $-1$ at $\beta t=4$, where we take 10000 Monte Carlo samples.
For comparison, we define a ratio of the CPU time of the rr-DMFT to that of the CDMFT and show the results in Fig.~\ref{fig:cputime}(b).
Even when the cluster symmetry is not used, the rr-DMFT reduces the computational cost, except for small $N_{c}$'s.
When we use the symmetry in the rr-DMFT, the computational cost becomes always smaller than that by the CDMFT including the case of small $N_{c}$.
Note that in the CDMFT we need to solve always the whole impurity problem of the $N_{c}$ site irrespective of the use of the cluster symmetry while in the rr-DMFT the use of the symmetry reduces the number of the impurity problems to be solved, yielding a large reduction of the total computational cost. 
Although the statistical error in the CDMFT can be reduced by the use of the symmetry, it is still larger than that in the rr-DMFT which solves smaller clusters.

Note that the slightly larger cost at $N_{c} \leq 16$ is ascribed to the relatively large number of invoking the $N_{s}$-site solver in the rr-DMFT at small $N_{c}/N_{s}$. 
For instance, when we solve a four-site cluster model with a two-site impurity solver, we need to use the solver 12 times and this is larger than $N_{c}^{3}/N_{s}^{3}=4^{3}/2^{3}=8$, the ratio of the computational cost of $N_{c}$- and $N_{s}$-site impurity solvers. 
The latter ratio increases more rapidly than the number using the solver as $N_c$ increases.

In a frustrated case ($t^{\prime}/t=-1$), a more practical comparison for the CT-AUX results can be made by normalizing the costs with respect to the average sign since the negative sign problem reduces the effective number of the Monte Carlo sampling.
The renormalized ratio of the CPU time is shown in Fig.~\ref{fig:cputime}(c), and the average sign is plotted in Fig.~\ref{fig:cputime}(d).
The rr-DMFT more efficiently reduces the CPU time for larger $N_c$, where the negative sign problem is severe for the $N_c$-site CDMFT while that in the rr-DMFT stays at the same level as that of $N_{s}=2$-site CDMFT simulations.

At the end of this subsection, we remark on the number of Monte Carlo samplings in our actual computation.
When the size of the quantum Monte Carlo solver is increased, it is necessary to take more samples to keep the magnitude of statistical errors of physical quantities at a certain level.
Namely, for a fixed number of samples, the $N_c$-site rr-DMFT gives a smaller statistical error than the $N_c$-site CDMFT since the former solves smaller cluster problems. 
For example, in the calculation of Fig.~\ref{fig:cputime}(b) for $t^{\prime}=0$, the relative standard errors $\delta D/D$ of the double occupancy $D=\sum_{i=1,\dots ,N_{c}}\langle n_{i\uparrow}n_{i\downarrow}\rangle /N_{c}$ are 0.0105 in the $4\times4$-site CDMFT and 0.00360 in the rr-DMFT (without the use of the cluster symmetry). 
In other words, in order to suppress the statistical error of $D$ to the same level of the rr-DMFT, the CDMFT requires about $(0.0105/0.00360)^2$ $\simeq$ $8.5$ times more samples (i.e., CPU time). 
This fact shows that the rr-DMFT is practically more efficient than the data presented in Fig.~\ref{fig:cputime}(b).

\subsection{Spin structure factor}\label{ssec:ssf}
\begin{figure}[tb]
  \begin{center}
   \includegraphics[width=0.8\columnwidth]{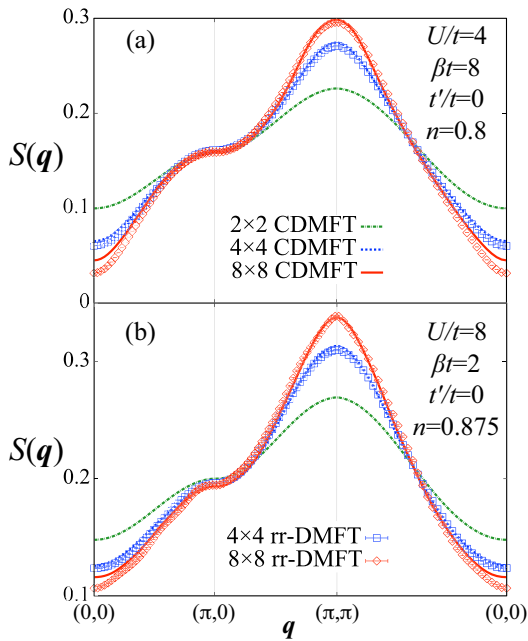}
  \end{center}
  \caption{ (Color online)
  Comparisons of spin structure factor $S(\bm{q})$ obtained by the rr-DMFT and CDMFT  for (a) $U/t=4, \beta t=8$ and $n=0.8$, and (b) $U/t=8, \beta t=2$ and $n=0.875$.
  Curves and symbols indicate the results of the CDMFT and rr-DMFT, respectively.
  An $N_{s}=4$-site impurity solver is used in the rr-DMFT.
  }
  \label{fig:ssfholedope}
\end{figure}

\begin{figure}[tb]
  \begin{center}
   \includegraphics[width=0.8\columnwidth]{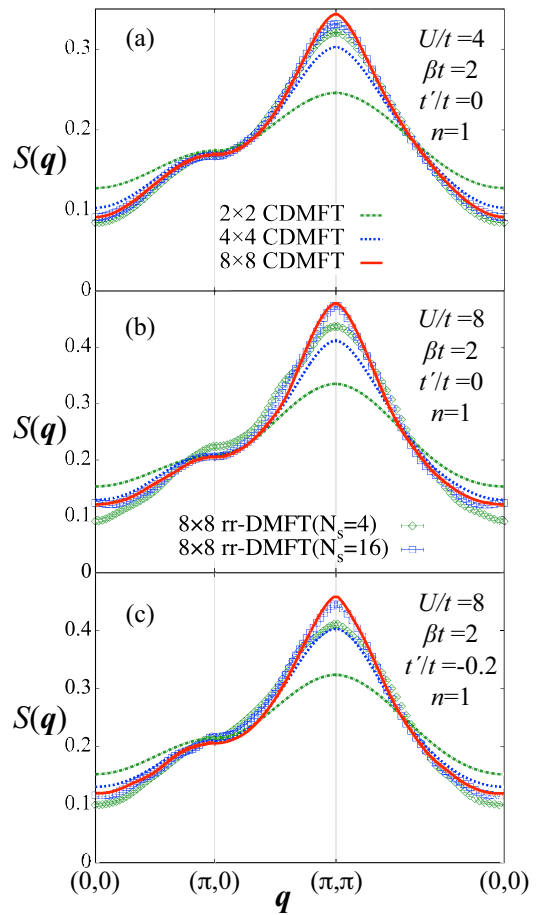}
  \end{center}
  \caption{ (Color online)
  Comparisons of spin structure factor $S(\bm{q})$ at half filling.
  The parameters are (a) $U/t=4, \beta t=2$ and $t^{\prime}/t=0$, (b) $U/t=8, \beta t=2$ and $t^{\prime}/t=0$, and (c) $U/t=8, \beta t=2$ and $t^{\prime}/t=-0.2$.
  Curves and symbols indicate the results of the CDMFT and rr-DMFT, respectively.
  We take $N_{s}=4$ and $16$ in the rr-DMFT simulations.
  }
  \label{fig:ssfhalffillingbeta2}
\end{figure}

In the two-dimensional Hubbard model, the short-range antiferromagnetic correlations play an important role in the Mott-insulating \cite{PhysRevB.76.045108, PhysRevLett.101.186403, doi:10.1143/JPSJ.76.084709} and pseudogap regions \cite{PhysRevB.66.075102, PhysRevLett.92.126401, PhysRevLett.95.106402, kuchinskii2005destruction, PhysRevB.73.165114, PhysRevB.74.125110, PhysRevB.76.104509, PhysRevB.80.165126,  PhysRevLett.102.056404, PhysRevB.82.155101, PhysRevB.84.075161, PhysRevB.82.134505, PhysRevLett.108.076401, PhysRevLett.86.139}.
The spin structure factor 
	\begin{eqnarray}
		S(\bm{q})&&=\frac{1}{N_{c}}\sum_{i,j=1}^{N_{c}}\langle S^{z}_{i}S^{z}_{j}\rangle e^{-i\bm{q}\cdot \left( \bm{r}_{i}-\bm{r}_{j}\right)},
	\end{eqnarray}
at $\bm{q}=(\pi ,\pi )$ is a useful measure of the antiferromagnetic correlations.
Here $\bm{r}_{i}$ is the position vector at the $i$-th site.

Figure \ref{fig:ssfholedope} compares $S(\bm{q})$ obtained by the CDMFT for $2\times 2$, $4\times 4$, and $8\times 8$ sites, and by the rr-DMFT for $N_c=4\times 4$ and $8\times 8$ sites with the $N_{s}=4$-site CT-AUX solver.
Here we examine the accuracy of the rr-DMFT in (a) an intermediate-coupling ($U/t=4$, $\beta t=8$ and $n=0.8$) and (b) a strong-coupling ($U/t=8$, $\beta t=2$ and $n=0.875$) cases, where $n$ is the electron density.
The rr-DMFT results agree well with those of the CDMFT (which is equivalent to the rr-DMFT with $N_{c}=N_{s}$) up to $8\times 8$ sites. In particular, the rr-DMFT reproduces the growth of $S(\bm{q}=(\pi ,\pi ))$ with the cluster size $N_{c}$.
This result shows that our method takes the growth of the antiferromagnetic correlations into account at a level similar to the $N_c$-site CDMFT.

The results of $S(\bm{q})$ at half filling ($n=1$) are shown in Fig.~\ref{fig:ssfhalffillingbeta2}.
Here, we calculate it in (a) an intermediate-coupling ($U/t=4$, $t^{\prime}=0$ and $\beta t=2$), (b) a strong-coupling ($U/t=8$, $t^{\prime}=0$ and $\beta t=2$) and (c) a frustrated ($U/t=8$, $\beta t=2$ and $t^{\prime}/t=-0.2$) cases. 
For the rr-DMFT, we fix $N_c$ at 8$\times$8 and compare the $N_s=4$ and 16 cases.
While the rr-DMFT with $N_{s}=4$ well reproduces $S(\bm{q})$ obtained by the CDMFT, there are visible differences between them.
The discrepancy is improved in the $N_s=16$ result since the result directly takes into account long-range vertex functions within the $N_{s}$-site cluster.

\begin{figure}[tb]
  \begin{center}
   \includegraphics[width=0.8\columnwidth]{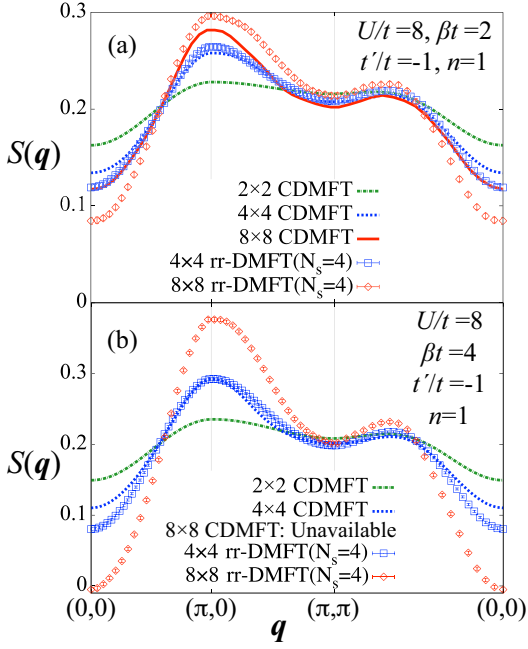}
  \end{center}
  \caption{ (Color online)
  Comparisons of spin structure factor $S(\bm{q})$ for a frustrated case ($U/t=8$ and $t^{\prime}/t=-1$) at half filling.
Here we compute $S(\bm{q})$ at (a) $\beta t=2$ and (b) $\beta t=4$.
Curves and symbols indicate the results of the CDMFT and rr-DMFT, respectively.
An $N_{s}=4$-site impurity solver is used in the rr-DMFT.
Note that at $\beta t=4$ the $8\times 8$-site CDMFT is not available because of its too large computational cost.
  }
  \label{fig:ssffullfrustrate}
\end{figure}

We also show a result for a frustrated case ($t^{\prime}/t=-1$) in Fig.~\ref{fig:ssffullfrustrate}.
In comparison with Figs.~\ref{fig:ssfholedope} and \ref{fig:ssfhalffillingbeta2}, the large frustration in Fig.~\ref{fig:ssffullfrustrate} suppresses the antiferromagnetic correlation, shifting the peak from $\bm{q}=(\pi ,\pi)$ to $(\pi ,0)$.
The rr-DMFT again reproduces well the overall spin structure of the $N_c$-site CDMFT.
The $S(q)$ peak at $\bm{q}=(\pi ,0)$ grows as $N_c$ increases in the CDMFT and it is also seen in the rr-DMFT.
Notice that at $\beta t=4$ [Fig.~\ref{fig:ssffullfrustrate}(b)] the 8$\times$8-site CDMFT is not available any more because of the severe negative sign problem [see Fig.~\ref{fig:cputime}(d)] while the 8$\times$8-site rr-DMFT is feasible and yields a result which looks reasonable, i.e., smoothly connected to the higher-temperature result [Fig.~\ref{fig:ssffullfrustrate}(a)].
Figure \ref{fig:ssffullfrustrate}(a) shows that the rr-DMFT slightly overestimates $S(\bm{q})$ for the $8\times 8$ cluster (basically within 10\% error). 
It is plausible that this originates from the increased number of coupling terms (next-nearest-neighbor hoppings $t^{\prime}$), which makes the convergence of the cumulant expansion slow, and leads to an underestimate of the spin frustration and resultant fluctuations. 
Nevertheless, Fig.~\ref{fig:ssffullfrustrate} demonstrates that the rr-DMFT reasonably works even when a severe negative sign problem makes the calculation of the standard CDMFT intractable.

\begin{figure}[!b]
  \begin{center}
   \includegraphics[width=0.8\columnwidth]{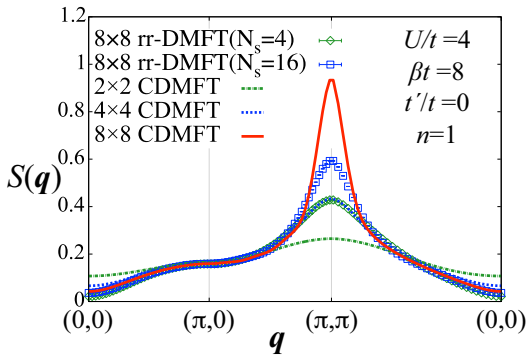}
  \end{center}
  \caption{ (Color online)
  Comparisons of spin structure factor $S(\bm{q})$ for $U/t=4$ and $\beta t=8$ at half filling.
  Curves and symbols indicate the results of the CDMFT and rr-DMFT, respectively.
  We take $N_{s}=4$ and $16$ in the rr-DMFT simulations.
  }
  \label{fig:ssfu4tn00beta8}
\end{figure}
Next, we show in Fig.~\ref{fig:ssfu4tn00beta8} low-temperature ($\beta t=8$) results for $U/t=4$ at half filling.
Comparing the rr-DMFT and CDMFT for the same $N_s$, we find that the rr-DMFT gives a larger value of $S(\bm{q}=(\pi ,\pi ))$, which is in fact closer to the value of the largest-cluster (8$\times$8) CDMFT result. However, when we compare the rr-DMFT and CDMFT for the same $N_c$, we see that the rr-DMFT underestimates the value of $S(\bm{q}=(\pi ,\pi ))$ while the increase of $N_s$ certainly improves the value.
This result indicates that, as is expected, the rr-DMFT is not sufficient when the system enters the critical region where the finite size effect is large or, in other words, the correlation length far exceeds $N_{s}$.
This should be attributed to the discard of the vertex corrections between decomposed small-size clusters in the rr-DMFT.

\subsection{Self-energy}\label{ssec:self energy}

\begin{figure}[tb]
  \begin{center}
   \includegraphics[width=\columnwidth]{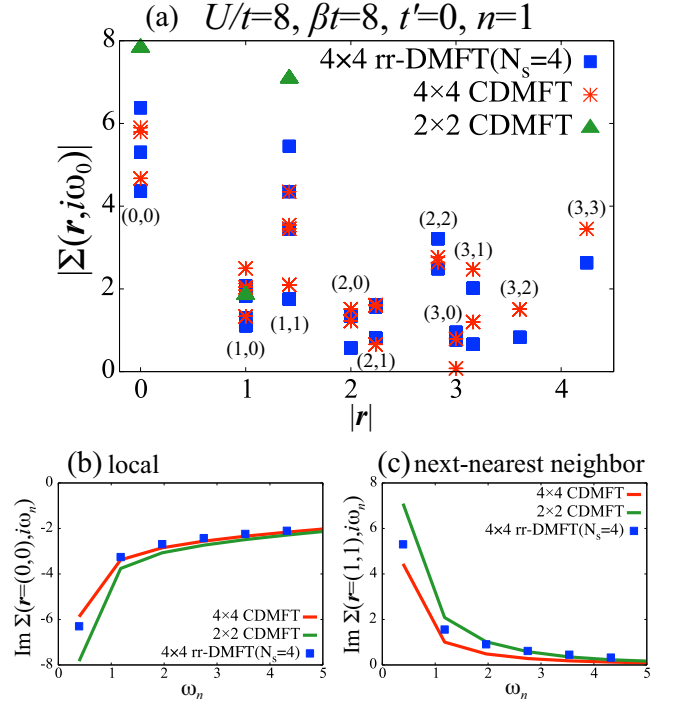}
  \end{center}
  \caption{(Color online)
  (a) Absolute value of self-energies, $|\Sigma (\bm{r},i\omega_{0})|$, obtained with rr-DMFT and CDMFT, at the lowest Matsubara frequency plotted against real-space distance $|\bm{r}|$ for $U/t=8, \beta t=8, t^{\prime}=0$, and $n=1$.
  The pair of numbers at each data represents $\bm{r}=(x,y)$ in unit of the lattice constant.
  There are several data at the same $\bm{r}$ since the methods break the translational symmetry.
  The imaginary parts of (b) the local and (c) next-nearest-neighbor self-energies at the central four sites in the $4\times 4$ cluster are plotted against the Matsubara frequency.
   }
  \label{fig:sigmaiw0}
\end{figure}

In this subsection, we examine the accuracy of the rr-DMFT for the self-energy in the real space.
We use the CT-AUX as the impurity solver.
The self-energy is an important quantity in the strongly-correlated systems since it determines the single-particle properties, such as the spectral weight.

While at high energy the self-energy in general decays with $1/\omega_n$ or even faster, it behaves differently at low energy, depending on the properties (metallic, insulating, and pseudogap etc.) of the system.
Therefore, as the most severe benchmark, we first examine the accuracy of the self-energy at the lowest Matsubara frequency $\omega_0$.
Figure \ref{fig:sigmaiw0}(a) plots the absolute value of $\Sigma (\bm{r},i\omega_{0})$ against the real-space distance $|\bm{r}|$.
We compare $2\times 2$- and $4\times 4$-site CDMFT and $4\times 4$-site rr-DMFT with $N_{s}=4$, for $U/t=8, \beta t=8, t^{\prime}=0$, and $n=1$.
Here several different values are obtained at the same $\bm{r}$ for each simulation, since the CDMFT and rr-DMFT intrinsically break the translational symmetry of the lattice.
For $|\bm{r}|<1.5$, we can see that the rr-DMFT gives the self-energy closer to the $4\times 4$-site CDMFT than to the $2\times 2$-site CDMFT. 
This is also seen in the $\omega_{n}$ dependence shown in Figs. \ref{fig:sigmaiw0}(b) and \ref{fig:sigmaiw0}(c). 
Moreover, Fig. \ref{fig:sigmaiw0}(a) shows that the rr-DMFT well reproduces the long-range part beyond the $2\times 2$ cluster at least up to $|\bm{r}|=3$. 
The lowest-order approximation in the cumulant expansion, Eq. (\ref{eq:renaction}), works well for this long-range part because the nonlocal one-body terms in the quantum impurity action decay with the distance $|\bm{r}|$. 
This also explains the deviation for $|\bm{r}|>3$, where the rr-DMFT always underestimates the correlations, interpolating the $N_{s}$-site CDMFT (giving zero self-energy for this long-range part) and $N_{c}$-site CDMFT. 
Namely, the truncation of the many-body hopping processes due to the lowest-order approximation in Eq. (\ref{eq:renaction}) underestimates the correlation between distant sites, hence giving an underestimated self-energy, while it certainly takes into account the spatial correlation beyond the $N_{s}$-site cluster.

\subsection{Mott metal-insulator transition}\label{ssec:mit}
The critical onsite interaction strength $U_{c}$ of the MIT is another criterion for understanding how much the spatial correlations are taken into account in the method, as the previous studies with the cluster extensions of DMFT (see, e.g., Ref.~[\onlinecite{PhysRevB.76.045108}]) have elucidated that, in the two-dimensional Hubbard model, the short-range antiferromagnetic correlations considerably reduce $U_{c}$.

We here investigate the $N_{c}$ dependence of $U_{c}$ within the rr-DMFT and the CDMFT at half filling.
Using the CT-AUX as the impurity solver, we calculate the local Green's function at $\tau =\beta /2$, which captures the characteristics of metals and insulators because
	\begin{eqnarray}
	-\beta G\left( \frac{\beta}{2}\right) =\frac{\beta}{2}\int d\omega \frac{\rho(\omega )}{\cosh{\left( \beta \omega \right)}},
	\end{eqnarray}
is approximately the average of the density of states $\rho(\omega)$ for $|\omega| \lesssim T$.
\begin{figure}[tb]
  \begin{center}
   \includegraphics[width=0.9\columnwidth]{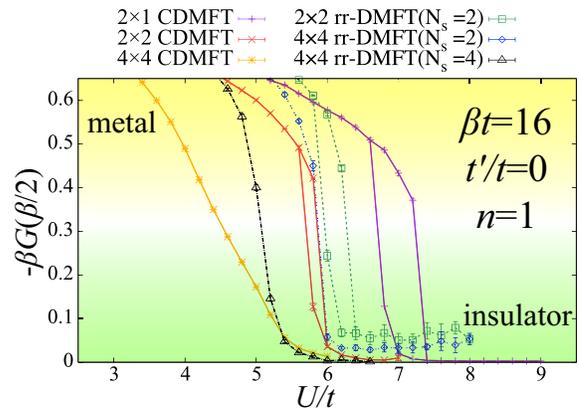}
  \end{center}
  \caption{(Color online)
  Comparison of $-\beta G(\beta /2)$ obtained by CDMFT (solid lines) and rr-DMFT (dashed lines).
   }
  \label{fig:mit}
\end{figure}
The results at $\beta t=16$ are shown in Fig.~\ref{fig:mit}.
For $N_{c}=2$ and 4, we find sudden changes of $-\beta G\left( \beta /2\right)$ with varying $U$, which are identified with the MIT.
In the CDMFT results, $U_{c}$ decreases from $N_{c}=2$ to 4, where the hysteresis proving the first-order transition \cite{PhysRevLett.101.186403} is observed.
The rr-DMFT quantitatively reproduces the reduction of $U_{c}$ with $N_c$ and the hysteresis for $N_{c}=4$.
The $2\times 2$-site rr-DMFT with $N_{s}=2$ indicates $U_{c}/t\simeq 6$, which is close to the $2\times 2$-site CDMFT result $U_{c}/t\simeq 5.8$.
This value is smaller than that of the $2\times 1$-site CDMFT, $U_{c}/t \sim 7.2$, despite the use of a two-site impurity solver in both calculations.
We find a crossover between a metal and the Mott insulator for the $N_{c}=16$-site CDMFT, since the hysteresis is not observed there.
It is plausible that the first-order MIT line ends at a temperature lower than $1/\beta=t/16$ for $N_c=16$.
The $4\times 4$ rr-DMFT with $N_{s}=2$ and $4$ reproduces this feature.

We conclude from these results that the rr-DMFT reasonably takes into account the spatial correlation beyond the size of the impurity solver.

\subsection{Density of states}\label{ssec:dos}
The ED solver \cite{PhysRevLett.72.1545} directly gives real-frequency quantities without use of any additional analytic continuation scheme, which is required in the CT-QMC.
As mentioned in Sec.~\ref{ssec:cost}, while the standard CDMFT with the ED solver requires a cost exponentially increasing with $N_{c}$, the rr-DMFT reduces the increase to a polynomial one.
Therefore, the combination of the rr-DMFT and the ED is a promising way to study the real-frequency properties of large systems.
In this subsection, we examine the accuracy of the rr-DMFT combined with the ED, by comparing the result with those obtained directly with the CDMFT.

\begin{figure}[tb]
  \begin{center}
   \includegraphics[width=0.9\columnwidth]{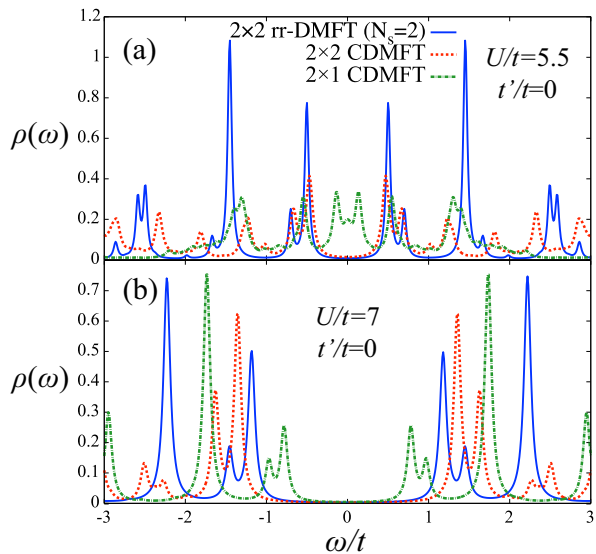}
  \end{center}
  \caption{(Color online)
  Density of states for (a) $U/t=5.5$ and (b) $U/t=7$ at half filling and zero temperature.
  Curves indicate the results of the $2\times 2$-site rr-DMFT (solid curves), $2\times 2$-site CDMFT (dashed curves) and $2\times 1$-site CDMFT (dot-dashed curves).
  Note that we use a broadening factor $\delta =0.05t$.
   }
  \label{fig:dos}
\end{figure}

Figure \ref{fig:dos} shows the density of states (DOS) $\rho (\omega )$ obtained by the CDMFT and the rr-DMFT with $N_{s}=2$-site ED impurity solver for $t'=0$ and half filling at zero temperature.
We take $n_b=2$ as the number of bath sites per impurity in all the simulations.
Note that we have introduced a broadening factor $\delta =0.05t$ as $\omega \rightarrow \omega +i\delta$ and shifted the chemical potential as $\mu \rightarrow \mu+U/2$ to fix the Fermi level at $\omega =0$.
At $U/t=5.5$, the $2\times 2$-site CDMFT has a gapped DOS and the rr-DMFT reproduces it qualitatively while the $2\times 1$-site CDMFT gives a metallic solution.
This result is consistent with the fact that the spatial correlation reduces $U_{c}$ as discussed in Sec.~\ref{ssec:mit}.
In Fig.~\ref{fig:dos}(b), all the solutions are gapped, where the rr-DMFT gives a gap size similar to that of the $2\times 2$-site CDMFT in comparison with that of the $2\times 1$-site CDMFT.
These results demonstrate that the rr-DMFT gives reasonable results even for the real-frequency properties, which are in general more sensitive to the numerical accuracy than the quantities defined on the Matsubara-frequency axis.

\section{CONCLUSION AND OUTLOOK}\label{sec:conclusion}
In summary, we have proposed the rr-DMFT as a method to efficiently take into account the spatial correlations at a computational cost substantially smaller than the brute force application of the CDMFT. 
We have applied the method to the two-dimensional Hubbard model on the square lattice.
We have demonstrated that the rr-DMFT substantially reduces the CPU time in comparison with ordinary CDMFT simulations.
Furthermore, when we use the CT-QMC as the impurity solver, the rr-DMFT substantially improves the negative-sign problem, leading to a further reduction of the CPU time.

As benchmark tests, we have demonstrated that the rr-DMFT well reproduces the spin structure factor, self-energy, the MIT point $U_{c}$ and the density of states, obtained by the CDMFT with the same cluster-size solver.

However, the rr-DMFT becomes insufficient when the antiferromagnetic correlation length is larger than $N_{s}$.
One possible prescription for this is, according to Ref.~[\onlinecite{PhysRevB.43.8549}, \onlinecite{PhysRevE.89.063301}], to take the higher-order cumulants into account in the real-space renormalization process, because it, in principle, partially includes the vertex corrections between the broken up clusters.
Of course, they include $n$-particle correlation functions defined by $\langle c^{\dagger}_{i_{1}}(\tau_{1})\cdots c^{\dagger}_{i_{n}}(\tau_{n})c_{j_{1}}(\tau_{1}^{\prime})c_{j_{n}}(\tau_{n}^{\prime})\rangle$ in the action Eq.~(\ref{eq:sc2}):
The leading correction to our method comes from the two-particle correlation functions.
This term will contain the spin-spin interaction which is related to the antiferromagnetic correlation.

Furthermore, while we have chosen the way of the cluster decomposition according to a physical insight (see Appendix B), it may be possible to apply an idea in the numerical linked-cluster expansion \cite{PhysRevLett.97.187202, PhysRevB.91.241107}, which fixes the way of the cluster decomposition according to a principle of inclusion and exclusion \cite{domb1974phase}, because both of them are based on a breakup of clusters into smaller blocks.

A potentially important application of the rr-DMFT is to calculate the real-frequency properties, such as the spectral function, for large clusters by using the ED solver, which does not rely on any additional analytic continuation scheme to obtain real-frequency properties.
Especially, our method will be useful to study the doped Mott insulators and frustrated systems.

It is also important to study multi-orbital models for the understanding of the role of the orbital degrees of freedom by extending the present rr-DMFT to multiorbital systems.

\section{ACKNOWLEDGMENTS}
The rr-DMFT code partly uses the nonequilibrium DMFT program developed by N. Tsuji in Ref. \onlinecite{RevModPhys.86.779}.
D. K. thanks T. Misawa, T. Ohgoe and Y. Yamaji, and S. S. thanks M. Ochi for helpful discussions.
Some of the results in this article have been computed at the facilities of the Supercomputer Center, Institute for Solid State Physics, University of Tokyo.
D. K. was financially supported by Japan Society for the Promotion of Science through Program for Leading Graduate Schools (ALPS).  
S. S. was supported by JSPS KAKENHI Grant No. 26800179.
This work was supported by the Computational Materials Science Initiative (CMSI), and RIKEN Advanced Institute for Computational Science through the HPCI System Research project (Projects No. hp140215 and No. hp150211).

\renewcommand{\theequation}{A.\arabic{equation}}
\setcounter{equation}{0}
\section*{APPENDIX A: Detailed algorithm}\label{sec:algorithm}
Here we describe the detailed algorithm, which solves an $N_{c}$-site quantum impurity model by using an $N_{s}$-site impurity solver $O\left( N_{c}/N_{s}\right)^{2}$ times.

First, we consider an $N_{c}$-site cluster and decompose it into four $N_{c}/4$-site blocks $C_{i} (i=1,\dots ,4)$ in accordance with Sec.~\ref{ssec:alg}.
We then categorize the elements of the Green's function into six groups: $A_{k}(k=1,2)$ and $B_{k}(k=1,\dots ,4)$.
In order to obtain each element, we calculate the action $S_{ij}$ which is given by tracing out the degrees of freedom other than $C_{i}$ and $C_{j}$ [for example, $(i,j)=(1,2)$ for $A_{1}$].
For $A_{1}$ and $A_{2}$, $G_{C_{i}C_{j}}$ and $G_{C_{j}C_{i}}$ are obtained from the actions $S_{12}$ and $S_{34}$, respectively, where $S_{ij}$ is the action consisting of $C_{i}$ and $C_{j}$ ($A$-type problem).
We here abbreviate a set $\{ G_{lm}| l\in C_{i}\wedge m\in C_{j} \}$ to $G_{C_{i}C_{j}}$.
For $B_{k}$ with $k=1, \dots ,4$, we calculate the elements of two sets $G_{C_{i}C_{j}}$ and $G_{C_{j}C_{i}}$ from the corresponding actions $S_{13}$, $S_{24}$, $S_{14}$ and $S_{23}$, respectively ($B$-type problem).
We thus calculate all the elements of $G_{ij}$ for the corresponding action $S_{ij}$ in the $A$-type problem, while we omit $G_{C_{i}C_{i}}$ and $G_{C_{j}C_{j}}$ in the $B$-type problem to avoid the overlap in the computation.

\begin{figure}[tb]
  \begin{center}
   \includegraphics[width=0.75\columnwidth]{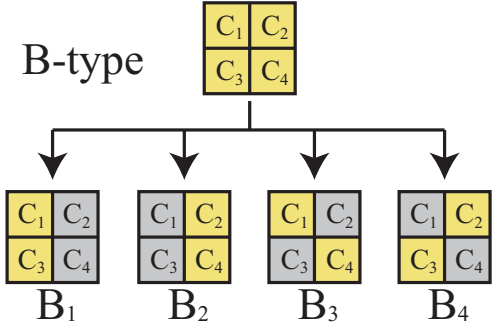}
  \end{center}
  \caption{(Color online) 
  Decomposition of a cluster and 
  resultant blocks which are used to calculate $B_{k} (k=1,\dots ,4)$ elements of Green's function.
  }
  \label{fig:rdmftalgorithmB}
\end{figure}
In the following, we evaluate the number using the $N_{s}$-site impurity solver for the $N_{c}$-site impurity problem.
We define $a(M,N_{s})$ and $b(M,N_{s})$ as the number applying the $N_{s}$-site impurity solver needed to solve the $M$-site $A$- and $B$-type problems, respectively.
The $A$-type problem is equivalent to an ordinary quantum impurity problem since we calculate all the elements of $G_{ij}$ within the cluster.
Because we are able to decompose a quantum impurity model into two $A$-type and four $B$-type problems, we obtain
	\begin{eqnarray}
	a(M,N_{s}) =&& 2a(\frac{M}{2},N_{s}) +4b(\frac{M}{2},N_{s})+\frac{9M}{4N_{s}}, 
	\label{eq:Atype0}
	\end{eqnarray}
where the last term on the right-hand side represents the number of times of using the $N_s$-site solver needed to reduce the $M$-site problem to the six $M/2$ site problems (namely, $A_1$ through $B_4$) as is proven below: 
We trace out the sites in $C=C_{1}+C_{2}+C_{3}+C_{4}$ in the following three independent steps:
(1) We first trace out $C_1$ and then $C_2$ or $C_4$, to obtain $S_{34}$ or $S_{23}$, respectively.
(2) We first trace out $C_2$ and then $C_3$ or $C_4$, to obtain $S_{13}$ or $S_{14}$, respectively.
(3) We first trace out $C_3$ and then $C_4$ or $C_1$, to obtain $S_{12}$ or $S_{24}$, respectively.
Thus, we eliminate the $M/4$-site block nine times with the $N_s$-site solver so that we have the last term in Eq.~(\ref{eq:Atype0}).
Note that, empirically, the order of tracing out does not significantly affect the results (see Appendix B).

Next, we consider the $B$-type problem for an $M$-site cluster.
We again decompose it into four problems (see Fig.~\ref{fig:rdmftalgorithmB}).
The elements $G_{ij}$ are classified into the following four parts;
	\begin{itemize}
	\item $\{ G_{C_{1}C_{3}}, G_{C_{3}C_{1}} \}$,
	\item $\{ G_{C_{2}C_{4}}, G_{C_{4}C_{2}} \}$,
	\item $\{ G_{C_{1}C_{4}}, G_{C_{4}C_{1}} \}$,
	\item $\{ G_{C_{2}C_{3}}, G_{C_{3}C_{2}} \}$.
	\end{itemize}
These are again $B$-type problems. 
Hence,
	\begin{eqnarray}
	b(M,N_{s}) =&& 4b(\frac{M}{2},N_{s}) +\frac{3M}{2N_{s}},
	\label{eq:Btype0}
	\end{eqnarray}
where $3M/2N_{s}$ comes from a reason similar to above. In this case, we trace out sites from $C=C_{1}+C_{2}+C_{3}+C_{4}$ as follows:
(1) We first trace out $C_1$ and then $C_2$ or $C_4$, to obtain $S_{34}$ or $S_{23}$, respectively.
(2) We first trace out $C_2$ and then $C_2$ or $C_4$, to obtain $S_{14}$ or $S_{13}$, respectively.

If $M=2^{n}N_{s}$ ($n>1$) is satisfied, the solution for $a$ and $b$ is obtained in the following way: Let us redefine 
$A_n=a(M,N_s)$ and $B_n=b(M,N_s)$, where we used the fact that $a$ and $b$ depend only on the ratio $M/N_s$.
From Eq.~(\ref{eq:Btype0}), the recursion relation for $B_{n}$ is obtained as 
	\begin{eqnarray}
	B_n +3\cdot 2^{n-1}=&& 4(B_{n-1} +3\cdot 2^{n-2}).
	\end{eqnarray}
Solving this recursion relation, since $B_2=b(2N_{s},N_{s})=8$ is easily shown, we obtain
	\begin{eqnarray}
	B_n=\frac{11}{4}2^{2n} -\frac{3}{2}2^n,
	\end{eqnarray}
resulting in
	\begin{eqnarray}
	b(N_{c}, N_{s})=\frac{11}{4}\left( \frac{N_{c}}{N_{s}}\right)^{2}-\frac{3N_{c}}{2N_{s}}.
	\end{eqnarray}
Similarly, using Eq.~(\ref{eq:Atype0}), the recursion relation for $A_{n}$ is obtained as
	\begin{eqnarray}
	A_n&-&\frac{11}{2}4^n+\frac{3n}{4}2^n \nonumber \\
&=& 2[A_{n-1} -\frac{11}{2}4^{n-1}+\frac{3(n-1)}{4}2^{n-1}],
	\end{eqnarray}	
The solution of the recursion relation leads to
	\begin{eqnarray}
	a(N_{c}, N_{s})=\frac{N_{c}}{4N_{s}} \left( \frac{22N_{c}}{N_{s}} -3 \log_{2}\left( \frac{N_{c}}{N_{s}}\right) -17 \right),
	\label{eq:Atyperesult}
	\end{eqnarray}
 where we have used the fact that $a(2N_{s},N_{s})=12$.
As a result, we find that the $N_{c}$-site impurity problem is solved by using the $N_{s}$-site impurity solver $O\left( N_{c}/N_{s}\right)^{2}$ times.
While Eq.~(\ref{eq:Atyperesult}) is valid only for $N_{c}=2^{n}N_{s}$ ($n>1$), the rr-DMFT algorithm itself is applicable to $N_{c}\ne 2^{n}N_{s}$ cases for which we expect a similar order of $a(N_{c},N_{s})$.

\renewcommand{\theequation}{B.\arabic{equation}}
\setcounter{equation}{0}
\section*{APPENDIX B: Tracing out order}\label{sec:traceorder}
Here, we elaborate our algorithm to determine the order of tracing out the sites. 
By comparing it with another algorithm, we show that the order does not significantly affect the final result. 
A basic principle of our algorithm is to choose a block of the sites to be traced out as compact as possible since the short-range correlation is better to be taken into account as much as possible.

\begin{figure}[tb]
  \begin{center}
   \includegraphics[width=\columnwidth]{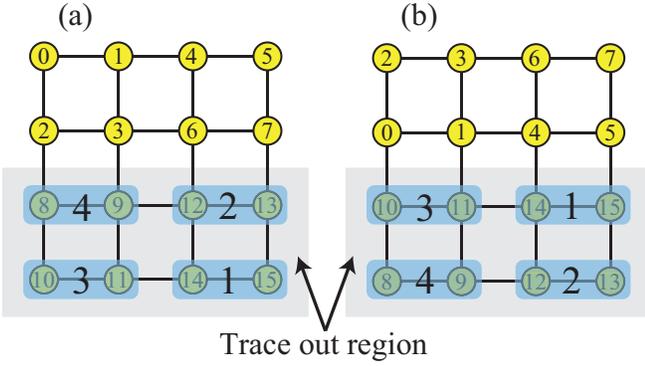}
  \end{center}
  \caption{(Color online) 
  (a) Site indices (indicated in yellow circles) and tracing out order (shown in blue shaded area) for $N_{c}=4\times 4$ and $N_{s}=2$.
 For the gray shaded sites, we trace out the blue shaded blocks in order of the number assigned to them.
  (b) 
  Different numbering of sites, to be used for the comparison made in Fig.~\ref{fig:szsztraceoutorder}.
  }
  \label{fig:traceoutorder}
\end{figure}
We consider a $2^{n}\times 2^{n}$ square cluster.
First, we assign numbers to all the sites according to the following rules:
(1) We divide the $2^{n}\times 2^{n}$ cluster into four blocks, $C_{i}(i=1,\dots ,4)$, along the cuts in the $x$ and $y$ directions as shown in Fig.~\ref{fig:rrdmftalgorithm1}, where $C_{i}$ is a $2^{n-1}\times 2^{n-1}$ cluster.
The upper left (right) block is $C_{1}$ ($C_{2}$) and the lower left (right) block is $C_{3}$ ($C_{4}$).
(2) For all the sites in the block $C_{i}$ $(i=1,\dots ,4)$, we assign $i-1$ to the $n$-th digit of the quaternary number representing the site.
(3) We iteratively apply the rules (1) and (2) with replacing $n$ with $n-1$ until $C_{i}$ is reduced to one site.
(4) After the loop of (1)$-$(3) finishes, we translate the obtained quaternary number into a decimal number to assign to each site.
An example of the $N_{c}=4\times 4$ case is shown in Fig.~\ref{fig:traceoutorder}(a).
Together with the breakup rule of a cluster described in Sec.~\ref{ssec:alg}, we trace out the sites in inverse order of the indices.
Figure \ref{fig:traceoutorder}(a) shows an example of tracing out a lower half part of a $4\times 4$ cluster with a two-site impurity solver.
When removing the sites in the gray region, we trace out the blue shaded two-site blocks in order of the indices assigned to them.
With this rule, blocks are automatically traced out in units of a block consisting of adjacent $N_{s}$ sites.
Moreover, the block $C_{i}$ always consists of adjacent sites too.

\begin{figure}[!b]
  \begin{center}
   \includegraphics[width=0.9\columnwidth]{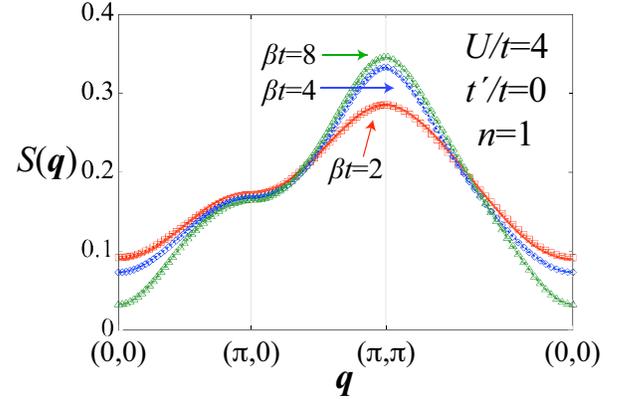}
  \end{center}
  \caption{(Color online) 
  Comparison of spin structure factors for two different tracing out orders shown in Fig.~\ref{fig:traceoutorder}.
  Curves and symbols respectively indicate the results of Fig.~\ref{fig:traceoutorder}(a) and \ref{fig:traceoutorder}(b).
   }
  \label{fig:szsztraceoutorder}
\end{figure}

Now we study how much the rr-DMFT results depend on the order of tracing out the sites and show numerically that the dependence is small. 
We consider a square cluster of $N_{c}=4\times 4$ and use the $N_{s}=2$ CT-AUX solver. 
In addition to the above-described numbering in Fig.~\ref{fig:traceoutorder}(a), we consider another numbering shown in Fig.~\ref{fig:traceoutorder}(b), whose sites are traced out in an order different from the former one.
In Fig.~\ref{fig:traceoutorder}(b), the order of performing traces in the lower half sites is illustrated.
The results of them are shown in Fig.~\ref{fig:szsztraceoutorder}, where we apply the rr-DMFT to the half-filled Hubbard model in two dimensions at $U/t = 4$ for several choices of temperature. 
We can see that the order of tracing out does not largely affect the results. 
We indeed find that these two patterns give similar results in the parameter regions studied in this paper.
\bibliography{paper}

\end{document}